\documentclass[aps,pre,twocolumn]{revtex4}
\usepackage{graphicx}
\usepackage{amsmath}
\usepackage{amssymb}
\usepackage{bm}
\usepackage{pifont}
\usepackage{color}
\usepackage[FIGTOPCAP]{subfigure}
\usepackage[latin1]{inputenc}

\begin{document}

\title{Collective dynamics in systems of active Brownian particles with dissipative interactions}

\author{Vladimir Lobaskin and Maksym Romenskyy}
\affiliation{School of Physics, Complex and Adaptive Systems Lab, University College Dublin, Belfield, Dublin 4, Ireland}

\date{\today}
\begin{abstract}
We use computer simulations to study the onset of collective motion in systems of interacting active particles. Our model is a
swarm of active Brownian particles with internal energy depot and interactions inspired by the dissipative particle dynamics method,
imposing pairwise friction force on the nearest neighbours. We study orientational ordering in a 2D system as a function of
energy influx rate and particle density. The model demonstrates a transition into the ordered state on increasing the particle
density and increasing the input power. Although both the alignment mechanism and the character of individual motion in our
model differ from those in the well-studied Vicsek model, it demonstrates identical statistical properties and phase behaviour.
\end{abstract}
\maketitle

\section{Introduction}

Dynamic self-organisation and, in particular, mechanisms of swarming behaviour of microorganisms, cells, and animals remain one
of the most intriguing problems at the interface of physics and biology. Numerous physical models of interacting self-propelled
particles have been proposed recently to study these phenomena (see review papers
\cite{toner.j:2005,ramaswami.s:2010,romanczuk.p:2012,vicsek.t:2012}). All these models capture the essential prerequisites for swarming:
out-of-equilibrium state, which is manifested in the self-propulsion of particles or other mechanisms of transforming
external energy into directed motion, and aligning or attractive interactions between the particles. The motion of individuals has
been described in the simplest case by particles moving with a constant speed and subjected to angular noise (the Vicsek model
\cite{vicsek.t:1995,czirok.a:1999,chate.h:2008}). More advanced presentations of active agents include friction, thrust force,
and noise, like, for example, the active Brownian particle model (ABP) \cite{ebeling.w:1999}, or even a very detailed mechanics of
cell or animal locomotion \cite{neilson.mp:2011,kabla.aj:2012}. The interactions required for the transition from individual to
collective dynamics have been introduced in a variety of ways. In the Vicsek model, the swarming results from the action of
a collision-type interaction that aligns the velocity of each actively moving particle in a big ensemble to the average local
velocity \cite{vicsek.t:1995,czirok.a:1999}. Alternatively, the particles' individual direction of motion can be coupled to the
mean orientation or position of the swarm \cite{czirok.a:1996,czirok.a:2000,ballerini.m.:2008,ginelli.f.:2010}.
In some implementations, the type of many-body interaction depends on the distance between neighbouring particles \cite{couzin.id:2002}.

The active motion and collective behaviour has also been observed in a number of synthetic systems including chemotactic colloidal particles
\cite{paxton.w:2004,paxton.w:2005,howse.j:2007,golestanian.r:2007,thakur.s:2011,taktikos.j:2012,theurkauff.i:2012},
Brownian machines and ratchets (see \cite{schweitzer.f:2007} for a review). Swimming particles with hydrodynamic interactions studied
theoretically using simulations or direct solution of the Stokes equation also showed an onset of
collective dynamics \cite{hernandez.jp:2005,llopis.i:2006}. Because of the microscopic size, their motion
is a subject to both passive and active fluctuations, which suggests that the ABP model, based on the Langevin equation for the velocity, could be
more appropriate in these cases than the models assuming constant propulsion speed. The ABP model has received much attention in
literature and has been successfully applied to a variety of problems \cite{romanczuk.p:2012}. Depending on the type of system under
study, different types of coupling between the ABPs have been used. Several realisations of the model assumed only conservative
\cite{ebeling.w:1999,ebeling.w:2000} or chemical interactions \cite{schweitzer.f:1994,schimansky-geier.l:1995} between moving agents.
Another development was based on the theory of canonical-dissipative system \cite{schweitzer.f:2001}. Lobaskin \emph{et al.} studied the Brownian dynamics
of a microswimmer and demonstrated its consistency with the ABP model \cite{lobaskin.v:2008}. Erdmann and Ebeling studied the active
Brownian particle model with Oseen-type hydrodynamic interactions \cite{erdmann.u:2003} and observed several swarming modes.
Recently, Grossmann et al. \cite{grossmann.r:2012} studied the onset of the collective motion
in a system of ABPs with velocity alignment and both passive and active fluctuations and found not only orientational
order-disorder transition but also bistable dynamics states.

The language of hydrodynamics is conceptually well suited for description of the swarm motion. This relation has been explored already in the
early papers by Toner and Tu \cite{toner.j:1995,toner.j:1998,toner.j:2005,ramaswami.s:2010}, who introduced a Navier-Stokes-like continuum
model for active materials. The hydrodynamic behaviour of active swarms can be inferred directly from microscopic description \cite{bertin.e:2009,ihle.t:2011}.
Bertin \emph{et al.} derived hydrodynamic equations governing the density and velocity fields from the microscopic dynamics for a gas of
self-propelled particles with pairwise interactions \cite{bertin.e:2009}. One can notice that similar ideas are
exploited in the mesoscale methods in fluid modelling such as multi-particle collision dynamics (MPCD)
\cite{malevanets.a:1999,malevanets.a:2000,gompper.g:2009}, where a collision operator is used to align particles to the average local flow
direction, or the dissipative particle dynamics (DPD) \cite{hoogerbrugge.pj:1992,koelman.jmva:1993}, where the
hydrodynamics comes in through inelastic collisions between the particles. Both of these methods designed to respect the momentum
transport (long-wave hydrodynamic modes) and to suppress the fluctuations (the high frequency modes) to achieve the
hydrodynamic behaviour at longer time and lengthscales. Obviously, the swarming behaviour can be achieved through MPCD or DPD-like
interactions as well. The possibility to develop the collective dynamics through dissipative interactions has been
recently investigated by Grossman et al. \cite{grossman.d:2008}. In their model, the system of active particles with spring-dashpot
interactions demonstrated a discontinuous transition into the aligned state upon reduction of noise and various types of
collective migration or vortex-like motion depending on the confinement. We are convinced that a system with dissipative interactions can have
at least qualitatively similar dynamics to the models with aligning interactions like the Vicsek model. In the spirit of these observations,
it is tempting to test whether the quantitative features of the swarming behaviour can be reproduced in a generic dissipative model upon
increasing energy influx.

In this paper, we study the dynamics of such a model and demonstrate that collective motion regimes can be achieved in the
same way as in the standard models of swarming, like the Vicsek model. We combine two well-developed approaches: the active Brownian particle
model, which allows us to introduce the self-propulsion and interactions with the environment in a transparent way, and dissipative
interactions for the active particles, so that the collective dynamics would arise from explicit pairwise forces. By analyzing
the statistical properties of this hybrid model, we show that it has the same universal properties across the order-disorder
transition as those reported for the Vicsek model. In Section \ref{Model} we
describe the construction of the model, in Section \ref{Results} we show its statistical properties in a wide range of parameters,
and calculate the phase boundary for the orientational order-disorder transition. We discuss the results in Section
\ref{Discussion}, and then conclude the paper in Section \ref{Conclusions}.

\section{Model and simulation settings}
\label{Model}

To study the dynamic self-organisation of active particles we introduce a two level model. At the single particle level we
include the factors determining the particle motion in a viscous medium: temperature and thermal
noise/fluctuations, friction, and a motor. At the two-body level, we introduce dissipative interparticle interactions. There 
we also include some noise, whose nature is, however, different from that of the environment. The noise at the many-body
level refers to biomimetic behavioural features like imperfect alignment of particles to their neighbours. Although the noise 
in the interaction is of non-thermal origin, it can also be characterized by some effective temperature. We will show that the 
characteristics of collective behaviour can be associated with these temperatures.

\subsection{Equation of motion of a single active particle}

At the single particle level, we follow the ABP model \cite{ebeling.w:1999}. Here, we consider the motion in two
dimensions. The motion of an individual particle $i$ is determined by the Langevin equation for the velocity with an added thrust
term
\begin{equation}\label{langevin}
     M\frac{d \mathbf{V}_i}{dt} = -\gamma^E \mathbf{V}_i + \sqrt{2 D^E} {\bm \xi}_i(t)+ \mathbf{F}_i^{T}.
\end{equation}
For simplicity, we will always set the particle mass $M$ to unity. The first term in Eq. (\ref{langevin}) is the standard Langevin
friction force. Here, $\gamma^E$ is the coefficient of viscous friction, which is set by the properties of the \emph{environment} and
the particle geometry, $\mathbf{V}_i$ is the velocity of particle $i$. Second term is a random force of strength $D^E$ and $\bm {\xi}(t)$
is representing Gaussian white noise with zero-mean and unit variance. The strength of the noise is set by the fluctuation-dissipation
relation at the ambient temperature $T^E$
\begin{equation}
D^E= \frac{T^E}{\gamma^E},
\label{sigma_e}
\end{equation}
where the temperature is expressed in energy units $k_B T$. In the following, we will use energy units for temperature. We should stress that the fluctuations introduced in Eq. (\ref{langevin})
act on each particle at all times and depend neither on the particle's speed nor on direction of motion.  Here, as we assume a fixed incoming power, this noise is determined by the
characteristics of the environment. This definition can, however, be generalized to include the fluctuations of the incoming power or thrust force.
Such situation can be realized in systems of chemically propelled particles \cite{thakur.s:2012}. If this power is normally
distributed, the behaviour of ABP will be qualitatively identical to the model with constant $q$ but with a redefined temperature
$T^E$ that would reflect the net amount of noise at the single-particle level \cite{sengupta.a:2011}.

\begin{figure}
\centering
\includegraphics[width=7.1cm,clip]{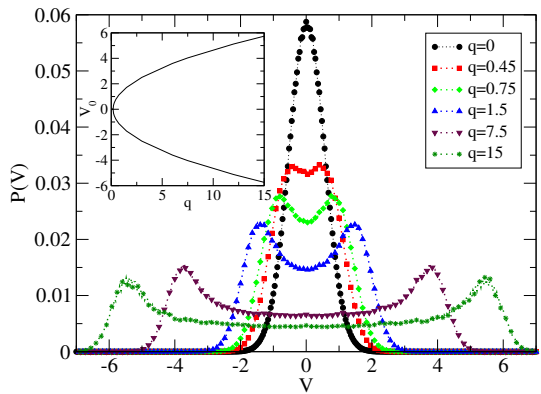}
\caption{(Color online) Instantaneous 1D velocity distributions for the active Brownian particles at $c=1.2$, $d=3.0$, $T^E=0.3$. \emph{Inset:}
Bifurcation diagram for the stationary velocity $V_0$ as a function of $q$.}
\label{fig:veld_q}
\end{figure}

The thrust term $\mathbf{F}_i^{T}$ (Eq. (\ref{langevin})) in the depot model has the form
\cite{schweitzer.f:2001}:
\begin{equation}
\mathbf{F}_i^{T} = \frac{q d}{c + d V_i^2} \mathbf{V}_i,
\label{nonlinear_friction}
\end{equation}
where $d$ is the constant determining the rate of conversion of internal energy of the active agent into kinetic energy, $c$ is
the parameter setting the internal energy dissipation rate, and $q$ is the constant determining the rate of energy influx
from the environment. The steady state motion of the active particles is characterized by velocity $V_0^2 = V_{0x}^2 + V_{0y}^2$, 
which is defined through the system's parameters as
\begin{equation}
V_0^2 = \frac{q}{\gamma^E} - \frac{c}{d}
\label{V0}
\end{equation}
at $q > \gamma^E c/d$ \cite{ebeling.w:1999,romanczuk.p:2012}. The steady state velocity distribution for various $q$ is shown in
Fig. \ref{fig:veld_q}. At $q=0$ we observe the Maxwell's distribution of the velocities corresponding to the system's
temperature $T^E$, while at non-zero energy influx rates we see either a broadened distribution centered at zero (at $q < \gamma^E
c/d$) or two bell-like peaks around the stationary velocity $V_{0x} = V_0/\sqrt{2}$ with the peak width controlled by the
temperature.
Thus, the model exhibits a transition from dissipative to driven regime upon increase of the energy influx rate $q$
\cite{ebeling.w:1999,ebeling.w:2000,erdmann.u:2003} and contains two well known limiting forms: Rayleigh (dissipative regime at low $q$)
and Schienbein-Gruler (driven regime at high $q$) \cite{romanczuk.p:2012}. Although the collective dynamics is observed only in the driven regime,
we use the most general expression for the thrust term to demonstrate the flexibility of the model. We also note that the phenomenology
of the order-disorder transition is not sensitive to the details of this term.

\subsection{Interparticle interactions}

The collective behaviour is impossible without interactions. While one can expect some swarming
(particle clustering) already with isotropic central interactions, the global symmetry breaking and onset of directed transport
requires that particle velocities are aligned. Here we should note that a spontaneous transition into a globally aligned state is
impossible in an equilibrium system with perfectly elastic collisions and without any dissipation due to conservation of the total
linear momentum. The local alignment can be realised by different means, the best known example being the Vicsek model, where
particles are aligned to the local mean velocity field \cite{vicsek.t:1995}. Another example of the aligning interaction
is the hydrodynamic interaction of fluid molecules, solute particles, or swimmers \cite{llopis.i:2006}. At the microscopic
level, the onset of hydrodynamic behaviour is achieved by suppressing the relative motion of the neighbouring particles with
a friction force and preserving the local mean velocity. As a result, the fluid quickly relaxes to the stationary state. This
idea is realised in a number of mesoscale simulation methods, which are known to produce correct hydrodynamics: lattice
Boltzmann (LB) method \cite{succi.s:2001}, MPCD, and DPD. In the LB and MPCD implementations, the collisions are collective, similar
to the Vicsek model, while in DPD the friction force is pairwise and is applied to each pair of colliding particles
\cite{pagonabarraga.i:1998}. So, the latter method is ideally suited for our purpose as it presents a simple way to control the
strength of the aligning interaction and relate it to other system's parameters.

Here, we introduce a dissipative force between the ABPs in the same way as it is done in the DPD method. The total force $\mathbf{F}_i(t)$ acting on each particle is then given by:
\begin{equation}
\mathbf{F}_i=\mathbf{F}_i^S -\gamma^E \mathbf{V}_i + \sqrt{2 D^E} {\bm \xi}_i(t)+ \mathbf{F}^{T}_i,
\label{total_force}
\end{equation}
where $\mathbf{F}_i^S$ is the force that comes from interactions within the swarm.  $\mathbf{F}_i^S$ consists of three parts:
\begin{equation}
{\mathbf{F}_i^S=\sum\limits^{}_{j\neq i}(\mathbf{F}_{ij}^{C}+\mathbf{F}_{ij}^{D}+\mathbf{F}_{ij}^{R})},
\label{DPD-force}
\end{equation}
where $\mathbf{F}_{ij}^C$, $\mathbf{F}_{ij}^D$, and $\mathbf{F}_{ij}^{R}$ represent the conservative, dissipative, and random forces  between particles
$i$ and $j$, respectively. The conservative force that reflects the excluded volume interactions is defined as:
\begin{equation}
{\mathbf{F}_{ij}^{C}=F^{C}(r_{ij})\hat{\mathbf{r}}_{ij}},
\label{cons_force}
\end{equation}
where $F^C(r)$ is a non-negative (repulsive) scalar function determining the distance dependence of the repulsion,
$\mathbf{r}_{ij}=\mathbf{r}_i-\mathbf{r}_j$ is the distance between particles $i$ and $j$, $r_{ij}= |\mathbf{r}_{ij} |$ is its
magnitude, and $\hat{\mathbf{r}}_{ij}=\mathbf{r}_{ij}/r_{ij}$ is the unit vector from $j$ to $i$ . We choose $F^C(r)$ to describe a
soft repulsion:
\begin{equation}
{F_{ij}^{C}(r)= \begin{cases} a \left ( 1-\displaystyle \frac{r}{r_r} \right ), &  r\leq r_r \\ 0, & r > r_r\end{cases}}.
\label{repulsive_f}
\end{equation}
Here, $a$ is a parameter determining the maximum repulsion force between the particles, $r_r$ is the radius of the repulsion zone. In the following, we will
assume $r_r=1$, so that the radius of repulsion, which can be interpreted as the body size of the active object, sets also a natural lengthscale of the problem.

The dissipative force suppresses the velocity differences between the neighbouring particles and, therefore,  provides a mechanism
of relaxation of the velocity field toward the stationary state. We take it in the form of a friction force applied to the
component of the motion in the direction of the particle connecting vector, i.e. a speed adjustment for particles moving
together in the same direction:
\begin{equation}
{\mathbf{F}_{ij}^{D}=-\gamma^S \omega^{D}(r_{ij})(\hat{\mathbf{r}}_{ij}\cdot\mathbf{V}_{ij})\hat{\mathbf{r}}_{ij}},
\label{dis_force}
\end{equation}
where $\mathbf{V}_{ij}=\mathbf{V}_i-\mathbf{V}_j$ is the relative velocity of particles $i$ and $j$.
Similarly, the friction can be applied to the motion perpendicular to the connecting vector \cite{jungans.c:2008},  in which case it
will predominantly act as an aligning interaction. In both cases, the parameter $\gamma^S$ controls the dissipative strength of the
interaction and by varying it we can accelerate or delay the alignment.

The non-conservative part of the DPD force can be used as a thermostat \cite{pagonabarraga.i:1998}. In this case, the stochastic force $\mathbf{F}_{ij}^{R}$ must be set to
compensate the loss of kinetic energy due to the dissipative force. It provides random "kicks'' in the radial direction $r_{ij}$ causing misalignment
of particle velocities.
\begin{equation}
{\mathbf{F}_{ij}^{R}(t)=\sqrt{2 D^S} \omega^{R}(r_{ij})\xi_{ij}(t)\hat{\mathbf{r}}_{ij}},
\label{stoch_force}
\end{equation}
where $D^S$ determines the strength of stochastic contribution to interactions, and $\xi_{ij}(t)$ is a random variable with a Gaussian
distribution and unit variance. In hydrodynamic simulations, it is usually required that the noise $\xi_{ij}$ is symmetric in $ij$,
the kicks satisfy Newton's third law and conserve total momentum \cite{espanol.p:1995}. This requirement, however, can be omitted
for active particles. The interactions can involve complex internal mechanisms of reorientation (like, for instance, contact inhibition
of locomotion - rearrangement of actin protrusions of motile cells \cite{neilson.mp:2011}), which do not conserve linear momentum.
We should stress that in this work we do not enforce the momentum conservation. Moreover, although we study the onset of collective
behaviour of the swarm, the interactions between the active particles are not mimicking the long-range hydrodynamic interaction of
microswimmers as can be represented, for example, by force dipoles \cite{erdmann.u:2003,mussler.m:2013}.

Despite the non-thermal nature of the interaction noise defined by Eq. (\ref{stoch_force}), we can define a swarm temperature, $T^S$, using the standard
fluctuation-dissipation relation
\begin{equation}
D^S =  \frac{T^{S}}{\gamma^{S}}.
\label{sigma}
\end{equation}
In Eqs. (\ref{dis_force})-(\ref{stoch_force}), $\omega^D(r)$ and $\omega^R(r)$ are weight functions addition of which lets us
ensure that the  fluctuation-dissipation relation holds \cite{espanol.p1:1995}. For simplicity $\omega^D(r)$ and $\omega^R(r)$ are defined as:
\begin{equation}
{\omega^{D}(r)=[\omega^{R}(r)]^2=\begin{cases} \left ( 1-\displaystyle \frac{r}{r_c} \right )^2, &  r\leq r_r \\ 0, &  r_r < r < r_c \end{cases}},
\label{weight_f}
\end{equation}
where $r_c$ is the interaction cut-off distance. In this model, we can regulate the interparticle interaction by changing the effective temperature, $T^S$, and the friction
coefficient, $\gamma^S$. This effective temperature determines the average degree of alignment the system can tolerate, while the
friction coefficient determines the dissipative strength of a single collision and the speed of relaxation toward the stationary state.
Note that in this case the friction and the noise depend on the particle relative position and velocities. Clearly, the global
ordering should depend on both types of fluctuations, individual (coming from the environment or energy influx) and pairwise, as given by Eq. (\ref{stoch_force}).
The whole set of the DPD terms thus reflects the behavioural contributions to the motion. For an animal or robotic systems it amounts to respecting the
excluded volume and adjusting the motion to the neighbours. The stochastic term in this context plays a role of angular noise or
errors of alignment of the agents to their neighbours' direction of motion.

\subsection{Simulation settings and motion statistics}

We used a two-dimensional system with periodic boundary conditions. The primary box size was fixed at $130 \times 130$
units and we varied the number of particles in the interval from 500 to 50000 to set the required density $\rho$. Simulations were performed with
time step of $\Delta t=0.005$. Particles were propagated using the Verlet algorithm \cite{verlet.l:1967}:
\begin{equation}
{\mathbf{r}_{i}(t+\Delta t)=2\mathbf{r}_{i}(t)-\mathbf{r}_{i}(t-\Delta t)+\Delta t^2\mathbf{F}_i(t)}.
\label{motion}
\end{equation}
The velocities of particles were calculated using St\"{o}rmer-Verlet method:
\begin{equation}
{\mathbf{V}_{i}(t)=\frac{\mathbf{r}_{i}(t+\Delta t)-\mathbf{r}_{i}(t-\Delta t)}{2\Delta t}}.
\label{velocity}
\end{equation}
Total number of time steps in each run was $1\times10^7$. The statistics was collected in the steady state and each characteristic
of motion was calculated by averaging over 5 independent runs. All simulations were performed with the following set of key
parameters: $r_r=1$, $r_c=2$, $a=1$, $d=3$, $c=1.2$, $\gamma^E=0.45$. Throughout the paper we also use $\gamma^S=1.5$,
$T^S=0$, $T^E=0.3$, except where noted otherwise. To set the unit of time in our simulations, we choose a unit speed $v=1$ such that
a particle moving at $V =v$ would make a distance $r_r$ per unit time (as in Fig. \ref{fig:veld_q}). This definition can be
reformulated in terms of kinetic energy: our simulation units are such that an active particle moving at a speed of one body radius
per unit time would have a kinetic energy $E = M V^2/2 = 1/2$. Therefore, a temperature $T^E=0.3$ in our settings means that
the root-mean-square speed of particles without propulsion ($q=0$) is
$V_{rms} = \sqrt{T^E/M}=0.548$, i.e. 0.548 body radii per unit time. Other parameters of the ABP-DPD model were chosen to make the
dynamic features around the order-disorder transition clearer. We note that qualitatively the dynamic behaviour of the swarms does
not change significantly in the wide range of parameters and the only essential requirements are the onset of the driven regime and
the presence of aligning interactions.

To characterise the collective motion in our model we use two different velocity correlation functions. The velocity
autocorrelation function is calculated as
\begin{equation}
C(t)= \frac {1} {N} \left \langle  \sum_{i = 1 }^N  \frac {\mathbf{V}_i(0)\cdot \mathbf{V}_i(t)} {|\mathbf{V}_i(0)| | \mathbf{V}_i(t)|}\right  \rangle,
\label{VACF}
\end{equation}
where $\langle \cdot \rangle$ stands for the ensemble average. The two-point velocity correlation function is calculated as
\begin{equation}
C_\parallel (r)=  \frac {1} {N (N-1)} \left \langle \sum_{i = 1 }^N \sum_{j \neq i }^N  \frac { \mathbf{V}_i(t)\cdot \mathbf{V}_j(t) } {|\mathbf{V}_i(t)| | \mathbf{V}_j(t)|} \right \rangle,
\label{CCF}
\end{equation}
where $i$ and $j$ label particles separated by distance $r=|\mathbf{r}_{ij}|$. With this definition, two particles with
parallel (antiparallel) velocities give a correlation of $+1$ ($-1$). The angular brackets denote the ensemble average. To
characterise the swarming behaviour of the particles we also perform a cluster analysis. Cluster in our model is defined as a
group of particles with a distance between neighbours smaller or equal to the cut-off radius $r_c$, therefore, particles
interacting directly or via neighbouring agents are included into one cluster. We calculate the number of clusters and mean cluster
size.

We characterise the orientational ordering by the polar order parameter, which quantifies the alignment of the particle motion to the average instantaneous velocity vector
\begin{equation}
\varphi(t) = \langle \cos \theta_i(t) \rangle =  \frac{1}{N} \sum_{i=1}^N \frac{\mathbf{V}_i(t) \cdot \langle \mathbf{V}(t) \rangle }{ | \mathbf{V}_i(t)| | \langle \mathbf{V}(t) \rangle |},
\label{nematic2}
\end{equation}
where $\theta_i$ is the angle between the velocity of particle $i$ and instantaneous average direction of motion of all agents. This order
parameter has been extensively used to describe the orientational ordering in various systems of self-propelled particles
\cite{vicsek.t:1995,chate.h2:2008}. It turns zero in the isotropic phase and finite positive values in the ordered phase, which makes
it easy to detect the transition.

To locate transition points precisely we also calculated the Binder cumulant \cite{binder.k:1981}
\begin{equation}
G_L = 1- \frac{\langle \varphi^4_L \rangle_t}{3\langle \varphi^2_L \rangle^2_t},
\label{binder}
\end{equation}
where $\langle \cdot \rangle_t$ stands for the time average and $L$ denotes the value calculated in a system of size $L$. The most
important property of the Binder cumulant is a very weak dependence on the system size so $G_L$ takes a universal value at
the critical point, which can be found as the intersection of all the curves $G_L$ obtained at different system sizes $L$
\cite{chate.h2:2008} at fixed density. To detect the transition points in $q-\rho$ plane precisely we plot three curves for
different $L$ at constant density and find the point where they cross each other. Then, we use those points to construct the phase
diagram.

\section{Results}
\label{Results}
\subsection{Collective motion}

\begin{figure}
\centering
\subfigure[]{
\includegraphics[width=5.5cm,clip]{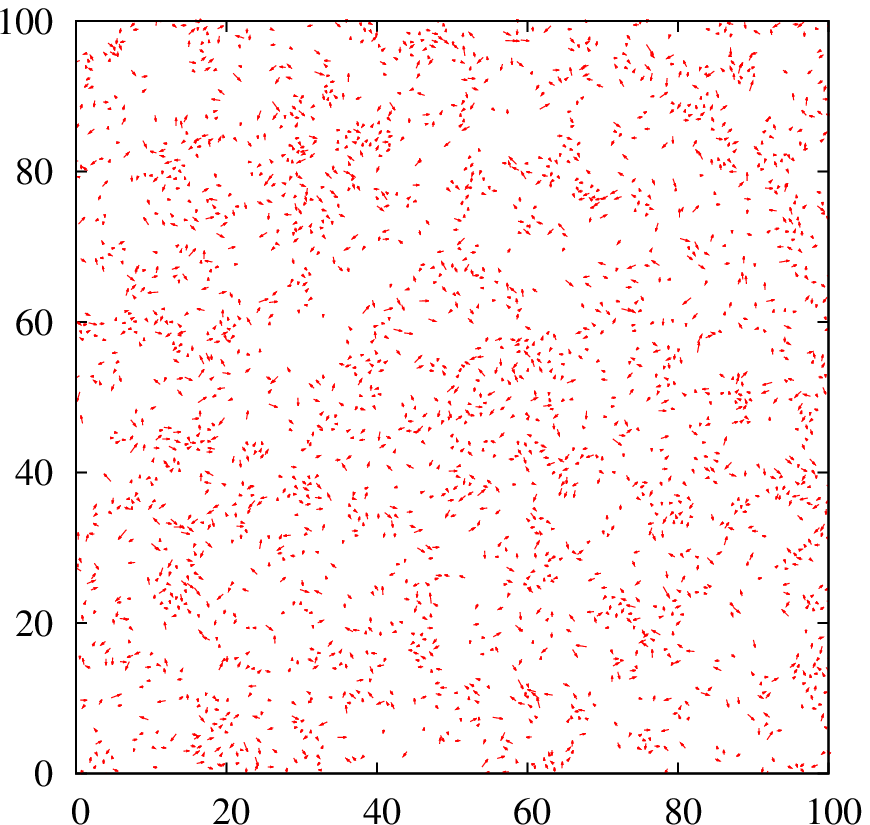}
}
\subfigure[]{
\includegraphics[width=5.5cm,clip]{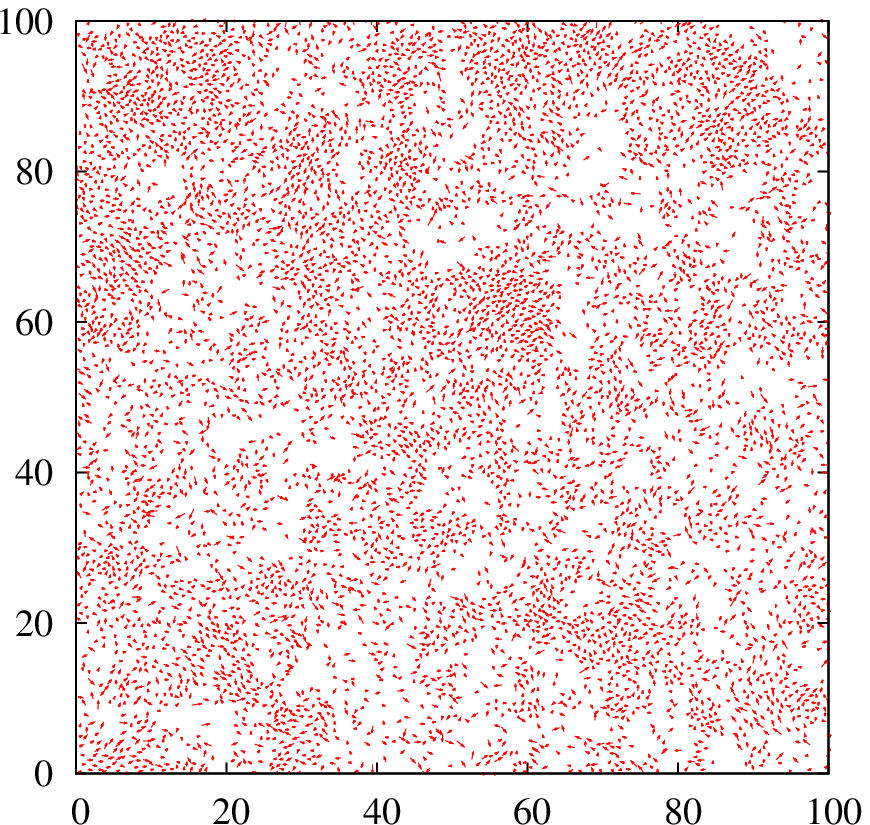}
}
\subfigure[]{
\includegraphics[width=5.5cm,clip]{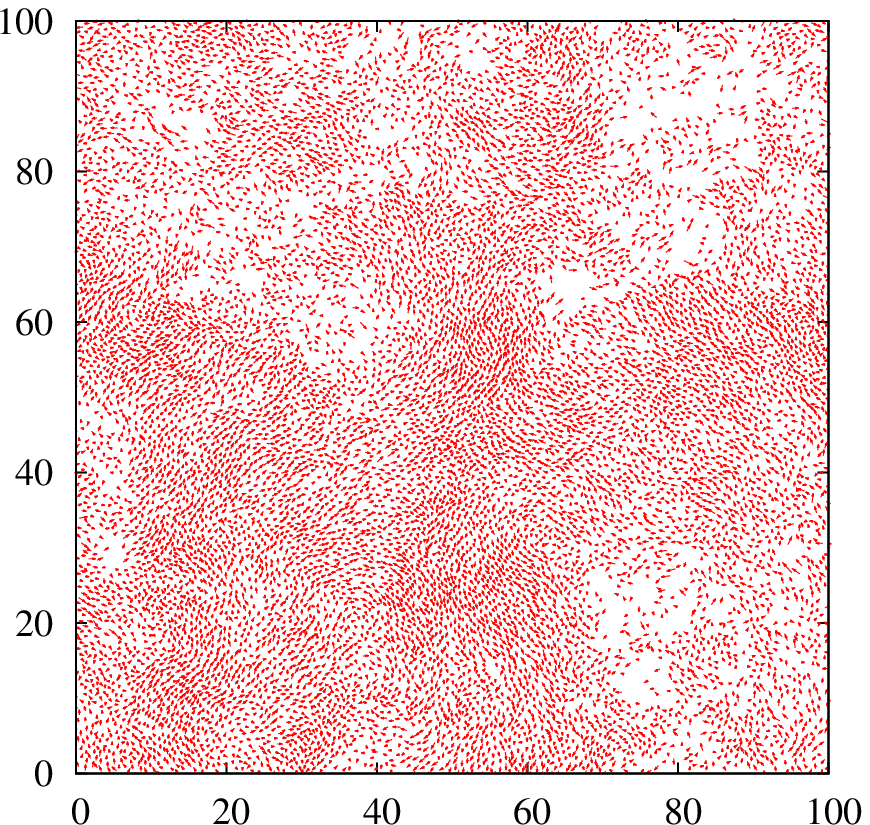}
}
\caption{(Color online) Typical distribution of particles inside a simulation box at constant propulsion power $q=0.45$ and different particle number densities: (a) $\rho=0.225$, (b) $\rho=0.675$, (c) $\rho=1.125$. Only part of the main box is shown for clarity purposes. Arrows indicate direction of motion of the individuals as well as velocity magnitude.}
\label{fig:snapshot_rho}
\end{figure}

We will illustrate the collective dynamics in our model by sequentially changing one of the two main parameters: the density
$\rho$ and the parameter controlling energy influx rate, $q$, which therefore determines the average propulsion speed of the
particles. In Fig. \ref{fig:snapshot_rho} we display simulation snapshots obtained at a fixed input power $q=0.45$ and different
particle number densities. There is no obvious global ordering in the system but we can detect formation of clusters. At fixed power 
$q$, an increase of the particle number density $\rho$ leads to stronger density fluctuations and the velocity alignment. 
In Fig. \ref{fig:snapshot_rho}(b),(c), we can notice a formation of dense particle groups, which move in the same direction. At low
density (Fig. \ref{fig:snapshot_rho}(a)), however, particles no large groups a seen and the particle velocities are oriented
randomly.

\begin{figure}
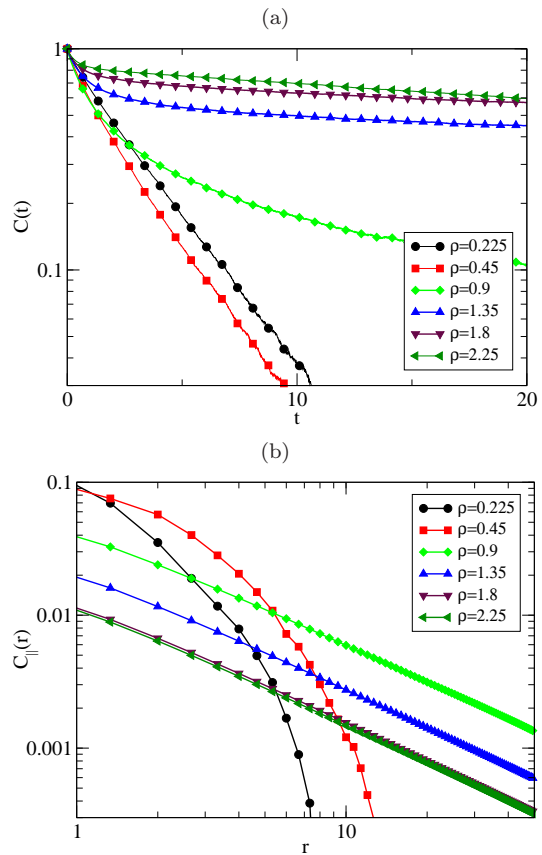

\centering
\subfigure[]{
\includegraphics[width=6.99cm,clip]{fig3a.eps}
}
\subfigure[]{
\includegraphics[width=7.0cm,clip]{fig3b.eps}
}
\caption{(Color online) Statistical properties of the ABP-DPD model at a constant energy influx rate $q=0.45$.
(a) Semi-log plot of velocity autocorrelation function $C(t)$ over time $t$.
(b) Spatial velocity correlation function $C_\parallel(r)$.}
\label{fig:param_rho}
\end{figure}

Figure \ref{fig:param_rho} illustrates the variation of statistical characteristics of the swarm upon a change of the
particle concentration. The velocity autocorrelation function, $C(t)$, in Fig. \ref{fig:param_rho}(a) shows an exponential decay
at low density, $\rho \leq 0.45$, with the decay time first decreasing  and then increasing with the concentration. 
At the higher densities, the decay changes dramatically, so the particles' direction of motion is getting much more stable in time. We should
note also that in the isotropic phase, $\rho = 0.225$ and $0.45$, the velocity correlation time is decreasing with concentration due
to the increase of the frequency of collisions. The trend is opposite in the ordered phase. This behaviour of the correlation time
also resembles the results for the Vicsek-type model, which we reported in Ref. \cite{romenskyy.m:2013}. The spatial velocity correlation
function, $C_\parallel (r)$, in Fig. \ref{fig:param_rho}(b) shows two distinct types of behaviour: the decay is exponential at the
two lowest concentrations, $\rho = 0.225$ and $0.45$, while it becomes algebraic at $\rho > 0.45$. We previously observed the
transition to the power law form for two-point velocity correlations for the Vicsek-type model \cite{romenskyy.m:2013}. 

\begin{figure}
\centering
\includegraphics[width=7.1cm,clip]{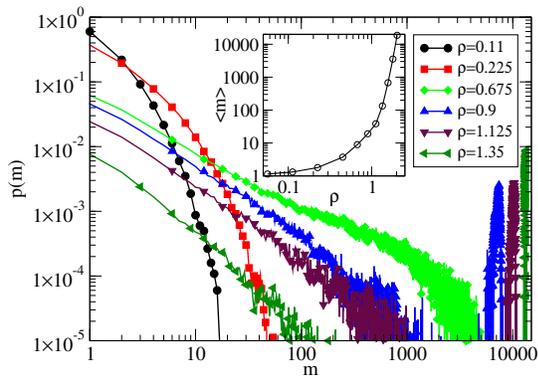}
\caption{(Color online) Cluster statistics for the ABP-DPD model at a constant energy influx rate $q=0.45$.
\emph{Inset:} The average cluster size. The exponent $p(m) \propto m^{-\zeta}$ for the straight segment: $\rho=0.675$
$\zeta\approx0.94$, $\rho=0.9$ $\zeta\approx$1, $\rho=1.125$ $\zeta\approx1.1$, $\rho=1.35$ $\zeta\approx1.3$.}
\label{fig:clstat_rho}
\end{figure}

The cluster statistics for the density series is shown in Fig. \ref{fig:clstat_rho}. The plot in the inset confirms our
observation that the cluster size is growing fast with the concentration. At $\rho=0.11$ and 0.225 the cluster size
distribution, as shown in the main plot, decays exponentially, while at larger density $\rho$ it changes into a power law, which
has been observed previously and is characteristic for the ordered phase \cite{huepe.c:2004,huepe.c:2008,romenskyy.m:2013}. Note that
at high densities, $\rho > 0.7$, majority of the particles belong to a single large cluster (narrow peaks seen at $m \approx 10000$)
and the relative weight of the small clusters is getting smaller. This trend is related to the growing overlap of the particle
alignment zones upon increase of the number density.


\begin{figure}
\centering
\subfigure[]{
\includegraphics[width=5.5cm,clip]{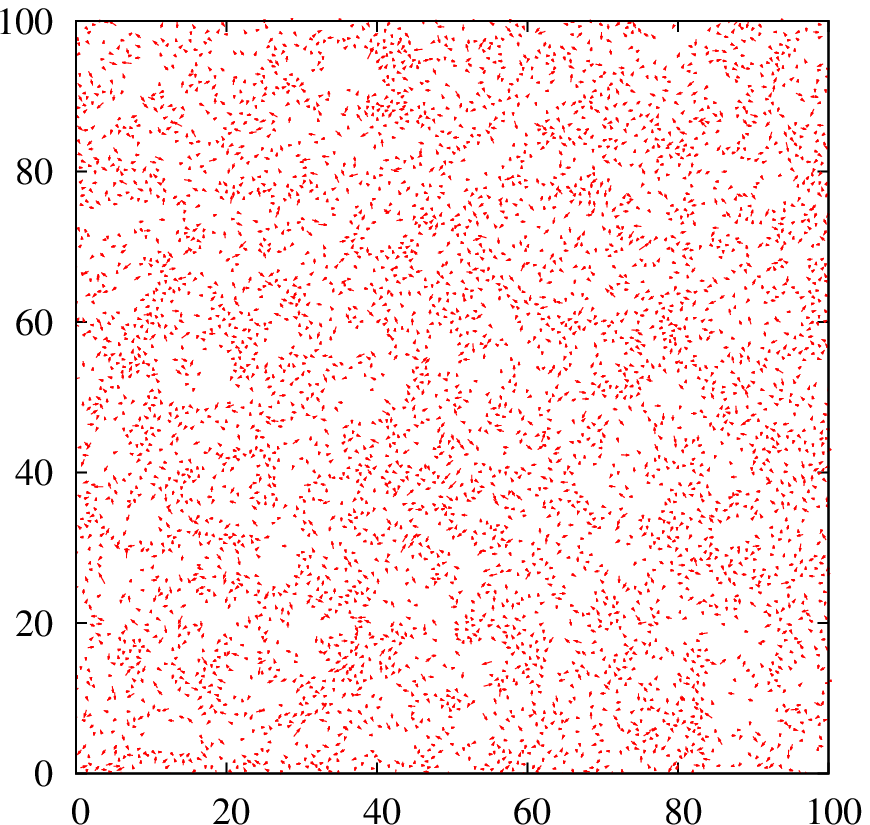}
}
\subfigure[]{
\includegraphics[width=5.5cm,clip]{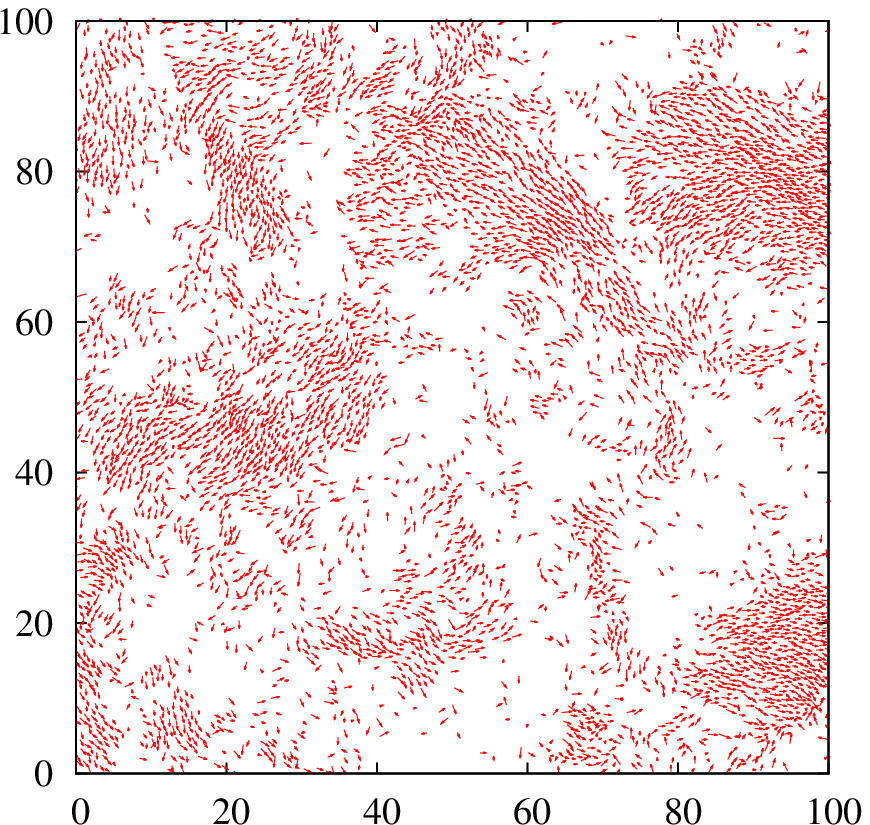}
}
\subfigure[]{
\includegraphics[width=5.5cm,clip]{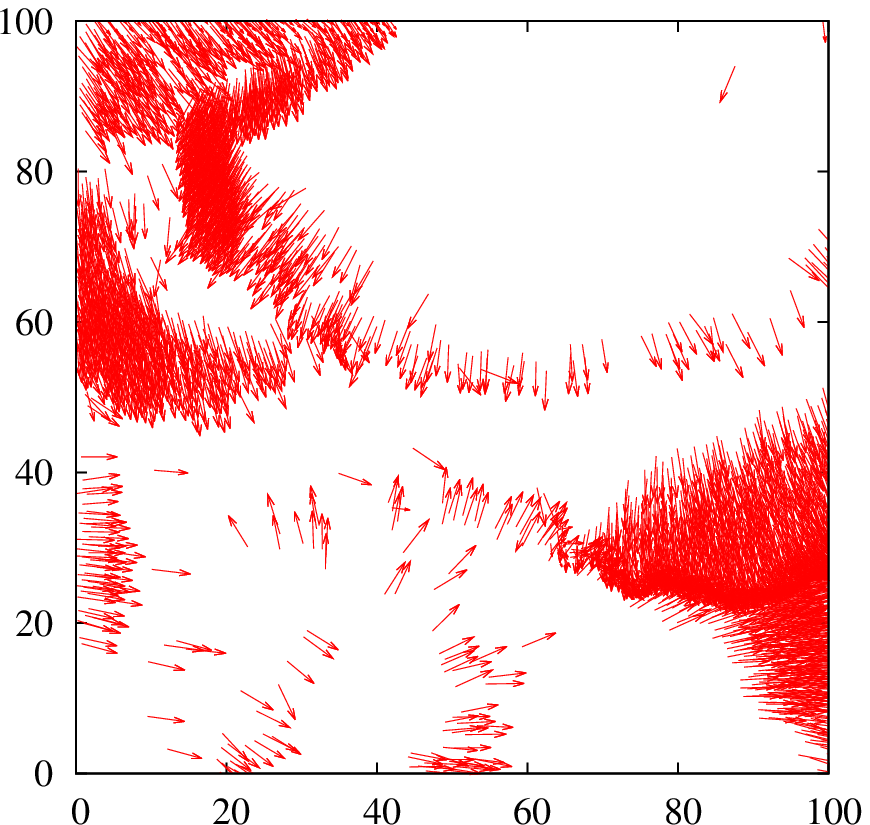}
}
\caption{(Color online) Typical distribution of particles inside simulation box at constant density $\rho=0.45$. (a) $q=0.3$, (b) $q=0.75$, (c) $q=15$. 
Only part of the main box is shown for clarity purposes. Arrows indicate direction of motion of the individuals as well as velocity magnitude.}
\label{fig:snapshot_q}
\end{figure}

Now, we will look at the behaviour of the system at constant density $\rho=0.45$ while varying the energy influx rate $q$. As
the ABPs change the behaviour from dissipative to driven upon increase of the energy influx rate, we expect the disordered
motion at low $q$ and onset of ordered behaviour at high $q$ levels.  Fig.~\ref{fig:snapshot_q} shows the alignment of
particle velocities at different input powers of the motor. Note that in the snapshots the arrows reflect the direction and the magnitude
of the instantaneous particle velocity, which varies with $q$. At low $q$ ($q=0.3$, Fig.~\ref{fig:snapshot_q}(a)) we see a
essentially homogeneous disordered system. Then, at $q=0.75$ (Fig.~\ref{fig:snapshot_q}(b)) distinct clusters are formed,
within which the particles move in nearly the same direction. At high $q$, Fig.~\ref{fig:snapshot_q}(c), the velocities are high,
the clusters are compact, and we observe a significant degree of alignment. In the system with $q=15$ there are very
few single particles and most particles belong to a single cluster. Note also the shape of the swarm: the ABP with dissipative interactions
tend to form bands, which are perpendicular to their velocity. It is important to mention that similar patterns have been observed
in other models of active particles. Thus, in binary mixtures of self-propelled particles stripe-like flocking behaviour arises from
inter-species interactions \cite{menzel.a:2012}. At certain density in Pursuit-Escape model \cite{romanczuk.p:2009} particles also
form clusters similar to ones observed in this study.

\begin{figure}
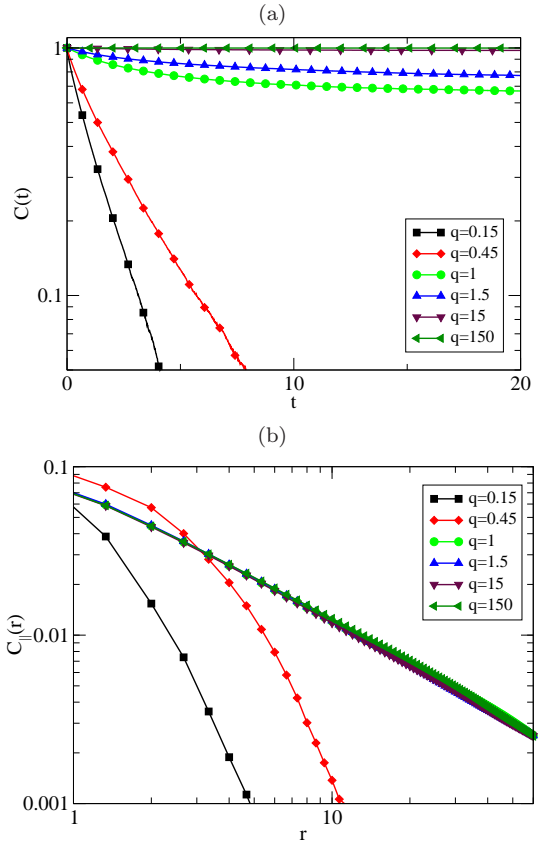

\centering
\subfigure[]{
\includegraphics[width=6.9cm,clip]{fig6a.eps}
}
\subfigure[]{
\includegraphics[width=7.1cm,clip]{fig6b.eps}
}
\caption{(Color online) Statistical properties of the ABP-DPD model at constant density $\rho=0.45$.
(a) Semi-log plot of velocity autocorrelation function $C(t)$ over time $t$. (b) Spatial velocity correlation function $C_\parallel(r)$.}
\label{fig:param_q}
\end{figure}

Analysis of the velocity autocorrelation function in our model at fixed concentration $\rho=0.45$ (Fig.~\ref{fig:param_q}(a)) shows
that at $q>0.45$ the motion of particles is very persistent. The change of the direction of motion is realised only through
collisions between different clusters. However, we see a fast decorrelation of the velocity at low $q$, $q=0.15$ and 0.45, due to
the thermal noise. In the spatial velocity correlations (Fig.~\ref{fig:param_q}(b)) we see a sharp transition from the
exponential decay at $q=0.15$ and 0.45 to a power law decay for larger $q$. All the curves showing the power law decay are
practically identical.

\begin{figure}
\centering
\includegraphics[width=7.1cm,clip]{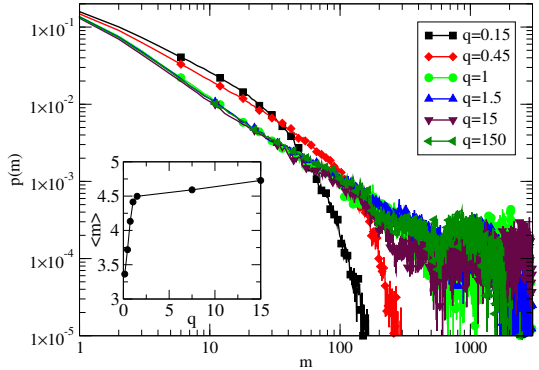}
\caption{(Color online) Cluster statistics for ABP-DPD model at constant density $\rho=0.45$. \emph{Inset:} Average cluster size.
The exponent $p(m) \propto m^{-\zeta}$ for the straight segment: $q=1$ $\zeta\approx1.16$, $q=1.5$ $\zeta\approx1.1$, $q=7.5$
$\zeta\approx1.1$, $q=15.0$ $\zeta\approx1.11$.}
\label{fig:clstat_q}
\end{figure}

Figure \ref{fig:clstat_q} presents the cluster statistics for particle concentration $\rho=0.45$. The main plot shows the cluster size
distribution on a log-log scale. We observe two qualitatively different distributions: at low energy influx rates, $q=0.15$ to
0.45, the curves show an exponential decay. At the higher $q$ all of them are practically identical and have a straight segment at
large numbers, which indicates the power law decay of the distribution. The transition can be located on the inset, where
the evolution of the mean cluster size is shown. We see a kink on the curve at $q \approx 0.6$. Note that no sharp peaks
corresponding to very large clusters appear in this figure as the particle number density is lower than that for some of the curves
shown in Fig. \ref{fig:clstat_rho}.

\begin{figure}
\centering
\includegraphics[width=7.0cm,clip]{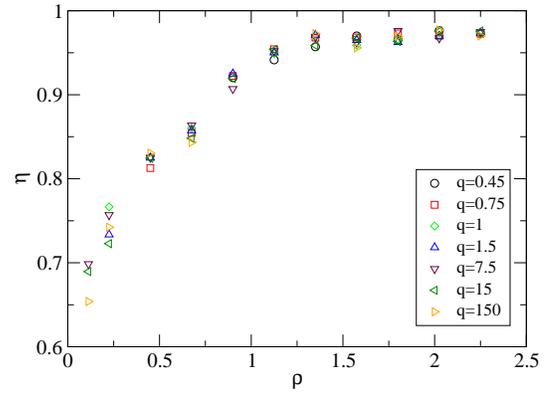}
\caption{(Color online) Behaviour of the exponent for the velocity correlation function, $C_\parallel(r)\propto r^{-d + 2 - \eta}$.}
\label{fig:power}
\end{figure}

We measured the decay exponent for the spatial velocity correlation function, $C_\parallel (r)$, in the whole range of
studied parameters $\rho$ and $q$ and found that the exponent assumes universal values that depend only on the density but not
on $q$, interaction parameters, or temperatures. The values of the exponent are plotted in Fig. \ref{fig:power}.

\subsection{Orientational ordering}

\begin{figure}
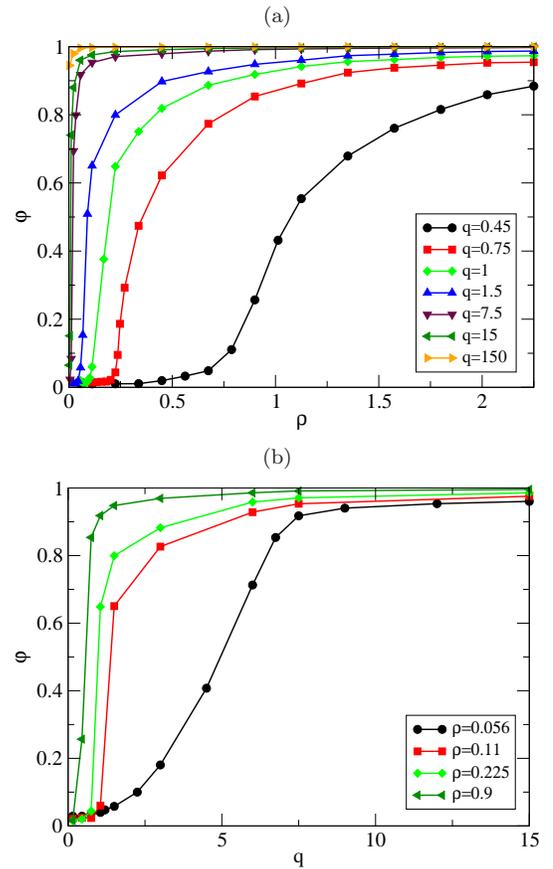

\subfigure[]{
\includegraphics[width=6.99cm,clip]{fig9a.eps}
}
\subfigure[]{
\includegraphics[width=7.0cm,clip]{fig9b.eps}
}
\caption{(Color online) Orientational order parameter for our model ($\varphi=1$ corresponds
to a completely ordered system, $\varphi=0$ - to a completely disordered system).}
\label{fig:order}
\end{figure}

Behaviour of the order parameter at various densities is shown in Fig. \ref{fig:order}a. At low densities the order parameter
values are close to zero during the whole simulation, which means that particle velocities are globally disaligned. At small
propulsive power, $q=0.45$, the ordering sets in slowly and reaches high values of about $\varphi \approx 0.8$ only at overlap
densities of $\rho \approx 2.25$. At very high driving power, $q=150$, the order parameter reaches unity at densities of about
$\rho = 0.09$ that corresponds to the mean distance between the particles $r = \rho^{-1/2} \approx 3.33$, which is greater than the
radius of interaction $r_c=2$. Obviously, the cohesive effect of the collisions keeps particles together, as can be seen already
from the snapshots in Fig. \ref{fig:snapshot_q}.

\begin{figure}
\includegraphics[width=6.95cm,clip]{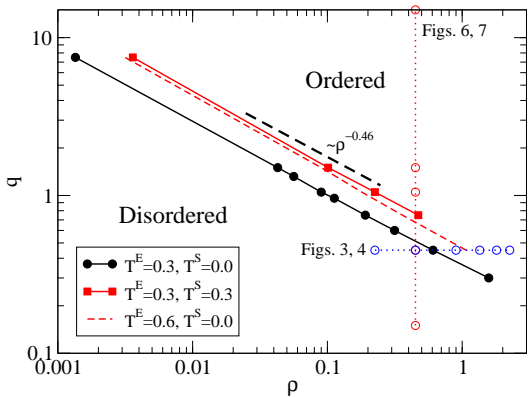}
\caption{(Color online) Phase diagram for the ABP-DPD model at $\gamma^E=0.45$, $\gamma^S=1.5$, $T^S=0$.
The blue and red open circles show the settings corresponding to series shown in Figs. \ref{fig:param_rho},
\ref{fig:clstat_rho} and \ref{fig:param_q}, \ref{fig:clstat_q}, respectively.}
\label{fig:phasediag}
\end{figure}

The phase diagram for our system is shown on Fig. \ref{fig:phasediag}. The region of ordered state corresponds to non-zero mean order 
parameter $\varphi>0$ while the disordered one to a vanishing particle mean velocity. The location of the transition points for
each set of parameters was determined using the standard Binder cumulant analysis from the intersection of the cumulant curves
$G_L$  calculated for three different system sizes. The behaviour of the cumulant indicates the continuous character of the
transition. It is clearly seen that the ordered behaviour, at fixed environmental noise, is possible at certain minimum energy influx
rate $q$, which, in its turn, determines the average propulsion speed. The critical energy influx rate changes with concentration
according to the power law $q_c \propto \rho^{-\kappa}$, where $\kappa$ is $0.46 \pm 0.02$. We show the transition lines for two
ambient temperatures, $T^E=0.3$ and $0.6$. The twice as higher temperature of the environment at fixed friction $\gamma^E$ means
that the passive fluctuations ($D^E$) are twice as more intense and a higher energy influx is required for the ABP to be able to
align. The $q_c$ values required for the transition at $T^E=0.6$ and $T^S=0$ are roughly 1.4 times higher than those found at
$T^E=0.3$ and $T^S=0$. The $q_c$ values observed $T^E=T^S=0.3$ are
very close to those obtained at $T^E=0.6$ and $T^S=0$.

\begin{figure}
\centering
\includegraphics[width=7.0cm,clip]{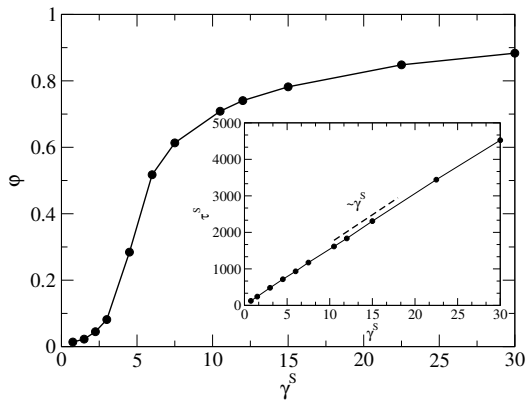}
\caption{Orientational order parameter in the ABP-DPD model as a function of
intraswarm friction $\gamma^S$. \emph{Inset:} Velocity correlation time vs friction $\gamma^S$
($T^S=0$, $\rho=0.45$, $q=0.45$).}
\label{fig:dif_gamma}
\end{figure}

Here, we would also like to demonstrate how the individual and collective dynamics of the particles depends on the key parameters
of the interaction. As mentioned above, the velocity correlations in our system decay exponentially in time in both phases according
to $C(t) \propto e^{-t/\tau^S}$ (see Fig. \ref{fig:dif_gamma}). We have measured the correlation time $\tau^S$ to demonstrate the
role of the intraswarm dissipation, which is controlled by $\gamma^S$. In a system without interactions, the relaxation time
would be completely determined by the dissipative and driving mechanisms of the Langevin equation and would normally decrease
with increasing the friction, $\tau^E =M/\gamma^E$. In contrast, as can be seen in the plot, the correlation time in the swarm,
$\tau^S$, is growing proportionally to $\gamma^S$. The friction coefficient $\gamma^S$ scales the dissipative power of the
pairwise collisions and therefore is the key parameter controlling the alignment. The pairwise friction acts only on relative motion
of the agents and therefore suppresses velocity fluctuations in the aligned state thus stabilising the motion. Indeed, in the main 
plot we see that the mean order parameter is also growing larger with $\gamma^S$.

\begin{figure}
\centering
\includegraphics[width=6.9cm,clip]{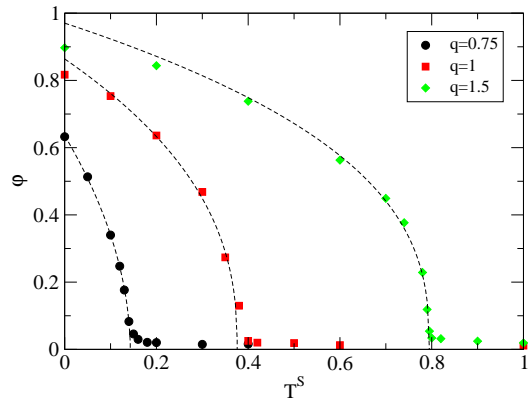}
\caption{(Color online) Behaviour of the order parameter as a function of the swarm temperature $T^S$
($\gamma^S=1.5$, $\rho=0.45$, $T^E=0.3$). The dashed lines show the power law fit to the points left of the transition temperatures.}
\label{fig:dif_T}
\end{figure}

Finally, Fig. \ref{fig:dif_T} illustrates the role of the swarm temperature $T^S$. The swarm temperature in our model can be
defined via a fluctuation-dissipation relation for the parameters of the pairwise interaction, noise, and the friction coefficient,
as given by the Eq. (\ref{sigma}). In terms of the temperature, the transition looks completely analogous to what is usually seen in the
magnetic systems. At zero temperature, the ordering is maximal, while it is suppressed by the fluctuations and vanishes at certain
maximal temperature $T^S_c$, which is getting higher at the higher input power $q$. The order parameter approaches zero according to a
power law $\phi \propto |T^S_c-T^S|^{\beta}$ with $\beta$=0.52 for $q=0.75$, $\beta=0.41$ for $q=1$, and $\beta=0.37$ for $q=1.5$,
which is in agreement with the critical exponent $\beta$ reported earlier for the Vicsek model and other models with aligning
interactions \cite{baglietto.g:2008,huepe.c:2008}.

\section{Discussion}
\label{Discussion}

As we can see from  the numerical data, a system of ABPs with dissipative interactions indeed demonstrates the same qualitative
properties as the well studied Vicsek model \cite{baglietto.g:2008,romenskyy.m:2013}. In contrast to most previous approaches, here we 
have separated two influences on the particle motion: the effect of the environment, which is introduced through the Langevin equation (\ref{langevin}), and the
effect of the other active agents, where the interactions are set by a separate pairwise dissipative parameter. Both effects can be
associated with a temperature, corresponding noise, and a friction parameter that control the rate of dissipation. Note
that these two types of noise and dissipation have different influence on the system. The former one is acting even on single
particles, while the latter applies only to pairs and vanishes for single agents. At zero temperature of
the environment, $T^E=0$, the model reduces to the motion with a constant speed, usually referred to as self-propelled particles,
as for instance in the Vicsek model. This regime would correspond in reality to a motion of macroscopic animals such that the thermal fluctuations
are negligible. At $T^S=0$, we have a system with aligning inelastic collisions but without the corresponding
noise in alignment. As we see from Fig. \ref{fig:dif_gamma}, the pairwise friction that scales the dissipation power in the
collisions can be used to regulate the degree of alignment in the system, which is expressed as the mean order parameter. Thus, our model
allows one to mix these contributions in different proportions and model different swarming scenarios.

Now, we would like to discuss the extent of the differences and similarities of the swarming behaviour in our model to observations from the Vicsek model in more detail. The main difference of the present analysis from the previous studies is that in our simulations we assumed a constant noise, as it is associated with the action of the environment, and followed the phenomenon as a function of the propulsive power of the particles. This path, however, can also be mapped onto a situation with a fixed particle speed and a variable noise. In static isolated systems, the ratio of the characteristic interaction energy to the thermal energy completely determines the equilibrium state. In the swarm of active particles, the crucial number is the ratio of the stationary velocity due to propulsion to the characteristic velocity due to thermal fluctuations. This ratio can be written as
\begin{equation}\label{ratio}
    \frac{V_0^2}{\langle V^2_{eq}\rangle  } \approx \frac{q}{\gamma^E} \frac{M}{T^E} = \frac{q \tau^E}{ T^E},
\end{equation}
where we used the relation between the friction and the relaxation time in the Langevin equation, $\tau^E = M/\gamma^E$, and the equipartition relation, $ M \langle V^2_{eq}\rangle = T^E$ . Thus, the ratio in question is equivalent to the incoming energy within the characteristic relaxation time, $q \tau^E$, to the thermal energy. In case the noise level is fixed by $T^E$, it is the stationary particle speed that matters. The mean speed in the ABP model at large $q$ is given by $V_0^2 = q/\gamma^{E}$. At higher temperatures of the environment, one needs to pump in more energy to produce the same ratio of the characteristic speeds. This point is confirmed by the data presented in Fig. \ref{fig:phasediag}. In case the stationary speed is fixed, one needs to reduce the temperature, which is, in the Langevin or DPD models, proportional to the fluctuation amplitude $D$. Therefore, the phase diagram in terms of noise amplitude $D$ vs density $\rho$ or $T^E$ vs $\rho$  will be inverse of our diagram shown in Fig. \ref{fig:phasediag}. It is interesting to note that the sum of the critical swarm temperature, as shown in Fig. \ref{fig:dif_T}, and the ambient temperature $T^E$ is roughly proportional to the energy influx rate. Here, we have $T_c^S+T^E \approx 0.2 + 0.3 =0.5$ for $q=0.75$, $T_c^S+T^E \approx 0.7$ for $q=1$, and $T_c^S+T^E \approx 0.8 + 0.3 =1.1$ for $q=1.5$. So, the fluctuations of different nature simply add up to increase the effective swarm's temperature, which can be defined as $T = T^E + T^S$ so that the ratio $q/T$ is about 1.5 in all cases. This idea is supported also by the phase diagrams shown in Fig. \ref{fig:phasediag}, where two systems with equal values of $T^E+T^S=0.6$ demonstrate a transition at nearly the same $q$ and $\rho$. In what regards the meaning of the temperatures entering this relation, we should note that although we assumed the noise in the Langevin equation strictly bound to ambient temperature, i.e. passive in nature, in some systems this term could be a combination of passive and active contributions so that the net magnitude of fluctuations corresponds to some effective temperature \cite{sengupta.a:2011}. Finally, we should also note that in this formulation the relation between the swarming phenomenon and spontaneous symmetry breaking in dissipative systems upon increase of the energy influx rate becomes more obvious \cite{prigogine.i:1977}. We hope to explore this relation in more detail in the future.

In simulations, we observe aggregation and orientational ordering of ABP at sufficiently high densities in presence of sufficiently high propulsive power. At the fixed level of noise and propulsive power, the cluster size grows with the particle concentration in the same way as we observed previously for the Vicsek model \cite{romenskyy.m:2013}. Thus, the dissipative interactions as well lead to cohesion of active particles. Secondly, they lead to particle alignment as can be seen from the growth of the order parameter with concentration, again, similar to the dependence seen in the Vicsek model. The transition into the orientationally ordered phase happens across the line $q_c \propto \rho^{-0.46}$, which is an inverse of the transition line for the Vicsek model, where it happens at $\xi \propto \rho^{0.45}$, where $\xi$ is the noise amplitude \cite{czirok.a:1997,romenskyy.m:2013}. This power law behaviour seems to be not unique to the Vicsek model, but a universal property of systems with global alignment and has been reported also for systems with pairwise aligning interactions (with an exponent $\kappa = 0.46 \pm 0.04$) \cite{bertin.e:2009}.

In what regards other properties, we should mention the behaviour of the correlations functions $C(t)$ and $C_\parallel (r)$ (Figs. \ref{fig:param_q} and \ref{fig:param_rho}), which demonstrate the same qualitative features as the Vicsek model we studied previously \cite{romenskyy.m:2013}. The two-point velocity correlation function changes the shape from exponential to a power law at the critical point and inside the whole region of the ordered behaviour. The exponent $\eta$, which describes the decay of $C_\parallel(r)\propto r^{-d+2-\eta}$ in the ordered phase, takes the same values from 0.5 to 0.97 on increasing density and shows the same density dependence as we previously saw in the Vicsek-type model \cite{romenskyy.m:2013}. It seems to be insensitive to other details of the system and reflects just the symmetry of the system. At the transition point, the exponent is expected to satisfy the Fisher's scaling law: $\gamma/\nu=2-\eta$ \cite{fisher.me:1974}, where $\gamma$ and $\nu$ are the critical exponents for isothermal susceptibility and the fluctuation correlation radius. In the limit of low concentrations, where the repulsions are not important, we have $\eta=0.5$, thus $2-\eta=1.5$, which is in agreement with the result for $\gamma/\nu=1.47$ obtained previously for the standard 2D Vicsek model \cite{baglietto.g:2008}. Moreover, the shape of the cluster size distributions as shown in Figs. \ref{fig:clstat_q} and \ref{fig:clstat_rho} in our model is also identical to that for the Vicsek model ranging from $\zeta=0.5$ to $\zeta=1.5$ depending on the level of noise and the density \cite{huepe.c:2008,huepe.c:2004,romenskyy.m:2013}. Although the type of active particle, the interactions and the type of noise differ from those in the Vicsek model, the identical values of the exponents suggest that our model belongs to the same universality class \cite{huepe.c:2008}.

\section{Conclusions}
\label{Conclusions}
We have studied dynamic self-organisation in a model combining the active Brownian particles with dissipative particle interactions, which are introduced via inelastic collisions. We found that the ABP-DPD model exhibits an orientational order-disorder transition on increasing energy influx rate or particle number density, which is completely  analogous to that in the Vicsek model, although the alignment mechanism in our model is completely different and the particle speeds are not constant. Moreover, the ABP-DPD system demonstrates the critical behaviour, which is identical to that of the Vicsek model. We have shown that the amount of ordering of such an active system can be characterized by effective temperatures of the environment and of the swarm and the ratio of the characteristic thermal energy to the energy influx per particle.

\section*{Acknowledgements}

Financial support from the Irish Research Council for Science, Engineering and Technology (IRCSET) is gratefully acknowledged.
The computing resources were provided by UCD and Ireland's High-Performance Computing Centre.

\vfil


\end{document}